\documentclass[aps,reprint,prb,twocolumn,showpacs,superscriptaddress]{revtex4-1}  
\usepackage{graphicx}  
\usepackage{dcolumn}   
\usepackage{bm}        
\usepackage{amssymb}   
\usepackage{amsmath}
\usepackage{color}
\usepackage[colorlinks=true,breaklinks=true,linkcolor=blue,urlcolor=blue,citecolor=blue]{hyperref}
\usepackage{gensymb}
\usepackage{bbm}
 
\begin{document}

\widetext

%

\title{The frustrated FCC antiferromagnet  Ba$_2$YOsO$_6$: structural characterization, magnetic properties and neutron scattering studies}

\author{E. Kermarrec}
\affiliation{Department of Physics and Astronomy, McMaster University, Hamilton, Ontario, L8S 4M1, Canada}
\author{C. A. Marjerrison}
\affiliation{Department of Physics and Astronomy, McMaster University, Hamilton, Ontario, L8S 4M1, Canada}
\author{C. M. Thompson}
\affiliation{Department of Chemistry, McMaster University, Hamilton, Ontario, L8S 4M1, Canada}
\author{D. D. Maharaj}
\affiliation{Department of Physics and Astronomy, McMaster University, Hamilton, Ontario, L8S 4M1, Canada}
\author{K. Levin}
\affiliation{Department of Chemistry, University of Manitoba, Winnipeg, Manitoba, R3T 2N2, Canada}
\author{S. Kroeker}
\affiliation{Department of Chemistry, University of Manitoba, Winnipeg, Manitoba, R3T 2N2, Canada}
\author{G. E. Granroth}
\affiliation{Neutron Data Analysis and Visualization Division, Oak Ridge National Laboratory, Oak Ridge, Tennessee 37831, USA}
\author{R. Flacau}
\affiliation{Canadian Neutron Beam Centre, AECL, Chalk River, Ontario, K0J 1J0, Canada  }
\author{Z. Yamani}
\affiliation{Canadian Neutron Beam Centre, AECL, Chalk River, Ontario, K0J 1J0, Canada  }
\author{J. E. Greedan}
\affiliation{Department of Chemistry, McMaster University, Hamilton, Ontario, L8S 4M1, Canada}
\author{B. D. Gaulin}
\affiliation{Department of Physics and Astronomy, McMaster University, Hamilton, Ontario, L8S 4M1, Canada}
\affiliation{Brockhouse Institute for Materials Research, Hamilton, Ontario, L8S 4M1, Canada}
\affiliation{Canadian Institute for Advanced Research, 180 Dundas St.\ W., Toronto, Ontario, M5G 1Z8, Canada}

\date{\today}

\begin{abstract}
We report the crystal structure, magnetization and neutron scattering measurements on the double perovskite Ba$_2$YOsO$_6$. The $Fm\overline{3}m$ space group is found both at 290~K and 3.5~K with cell constants $a_0 = 8.3541(4)$~{\AA} and $8.3435(4)$~\AA, respectively. Os$^{5+}$ ($5d^3$) ions occupy a non-distorted, geometrically frustrated face-centered-cubic (FCC) lattice. A Curie-Weiss temperature $\theta = -772$~K suggests the presence of a large antiferromagnetic interaction and a high degree of magnetic frustration. A magnetic transition to long range antiferromagnetic order, consistent with a Type I FCC state below $T_{\rm N} \sim 69$~K, is revealed by magnetization, Fisher heat capacity and elastic neutron scattering, with an ordered moment of 1.65(6)~$\mu_B$ on Os$^{5+}$.  The ordered moment is much reduced from either the expected spin only value of $\sim 3 \mu_B$ or the value appropriate to $4d^3$ Ru$^{5+}$ in isostructural Ba$_2$YRuO$_6$ of 2.2(1)~$\mu_B$, suggesting a role for spin orbit coupling (SOC). Triple axis neutron scattering measurements of the order parameter suggest an additional first-order transition at $T = 67.45$~K, and the existence of a second ordered state.
Time-of-flight inelastic neutron results reveal a large spin gap $\Delta \sim 17$~meV, unexpected for an orbitally quenched,  $d^3$ electronic configuration. We discuss this in the context of the $\sim 5$~meV spin gap observed in the related Ru$^{5+}$, $4d^3$ cubic double perovskite Ba$_2$YRuO$_6$, and attribute the $\sim 3$ times larger gap to stronger SOC present in this heavier, $5d$, osmate system.  
\end{abstract}


\maketitle

\section{Introduction}
%
Geometrical frustration in magnetic materials has proven to be extraordinary fertile ground for the discovery of new exotic quantum states of matter.\cite{IFM} Studies of materials with geometrically frustrated geometries such as the triangular,\cite{Shimizu2003,Olariu2006,Mourigal2014} kagom\'e,\cite{Mendels2007,Han2012,Fak2012} hyperkagom\'e\cite{Okamoto2007,Dalmas2003} or pyrochlore\cite{Gardner2010,Ross2011} lattices have yielded a host of topical disordered and ordered states, among them spin ice, spin liquid, and spin glass states for example. However, much less attention has been devoted to the face-centered-cubic lattice, comprised of edge sharing tetrahedra, which is also frustrated in the presence of antiferromagnetic interactions.

\begin{figure}
\includegraphics[width=\columnwidth]{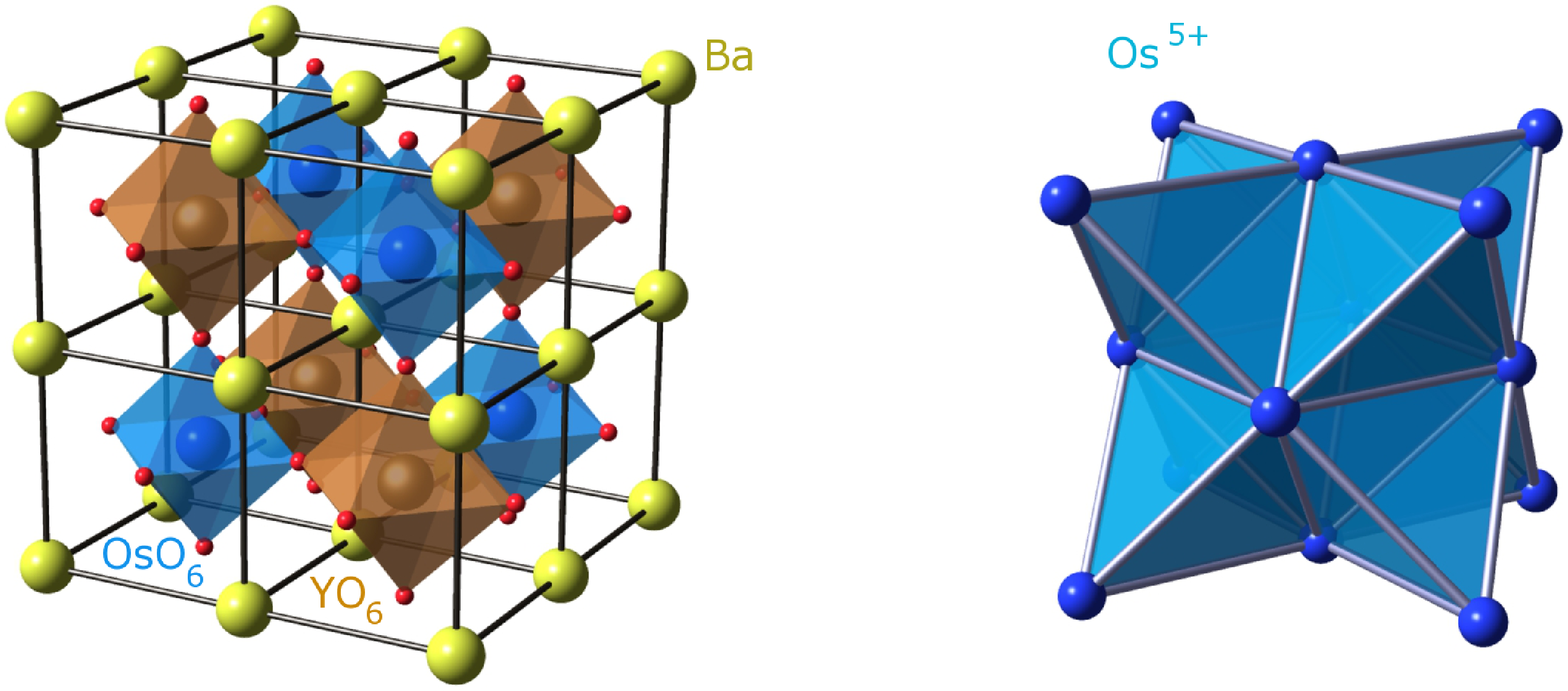}
\caption{\label{fig_stucture} (Color online) Left: Representation of the double perovskite structure of Ba$_2$YOsO$_6$ emphasizing the octahedral environment of Os$^{5+}$. Right: The magnetic Os$^{5+}$ ions form a frustrated FCC sublattice made of edge-sharing tetrahedra.}
\end{figure}

Such a lattice can be conveniently realized in the double perovskite structure, with chemical formula $A_2BB'O_6$. If the two ions $B$ and $B'$ order over the octahedral sites, two interpenetrating FCC lattices are formed and, if only $B'$ is magnetic, an FCC magnetic lattice results, as shown in Fig.\ref{fig_stucture}. Such a lattice is stabilized when the size and charge of the two cations $B'$ and $B$ are sufficiently different.\cite{Anderson1993} In general such materials can be prepared with very low levels of $B/B'$ intersite mixing. For example in the series Ba$_2$Y$M$O$_6$ where $M=$~Mo, Re and Ru, site mixing levels as determined by $^{89}$Y~NMR were reported as 3\%, $<0.5$\% and 1\%, respectively.\cite{Aharen2009,Aharen2010b,Aharen2010}

Despite possessing a relatively common structure, frustrated magnetic materials with FCC lattices have yet to be widely explored. The study of the double perovskite family is therefore a promising avenue in this regard, as shown by recent measurements underlining the diversity of stabilized states. The flexibility of the cubic double perovskite structure allows systematic investigation of a large fraction of the transition elements in the periodic table, going from those possessing rather large, classical, spin-only moments such as $S=\frac{3}{2}$ to the extreme quantum case of $S=\frac{1}{2}$. Of particular interest is the Ba$_2$Y$B'$O$_6$ family, where $B'$ is a magnetic $4d$ or $5d$ transition-metal element in its 5+ oxidation state. These materials tend to retain the ideal cubic $Fm\overline{3}m$ structure to the lowest temperatures studied. Examples include Ba$_2$YRuO$_6$ ($4d^3$),\cite{Aharen2009,Carlo2013}  which enters a long-range ordered antiferromagnetic state at $T_{\rm N} = 36$~K, well below its Curie-Weiss temperature $\theta_{\rm CW} = -571$~K, Ba$_2$YReO$_6$ ($5d^2$) which shows anomalous spin freezing behaviour below $T \sim 35$~K,\cite{Aharen2010} and Ba$_2$YMoO$_6$ ($4d^1$) which possesses a collective spin-singlet ground state.\cite{DeVries2010,DeVries2013,Carlo2011,Tarzia2008,Singh2010}

The double perovskite structure also allows for the study of magnetism of heavier $5d$ elements, where the spin-orbit coupling and the crystal field energy are known to become of comparable value. The originality of $5d$ insulating systems has been recently highlighted in the $5d^5$ iridate compounds, where a strong SOC induces a Mott instability leading to an effective total angular momentum $J_{\rm eff} = \frac{1}{2}$ state,\cite{Kim2009} very different from the well-known spin $S = \frac{1}{2}$ localized state of conventional Mott insulators.

The role of SOC for t$_{2g}^3$ ions with $d^3$ configurations, such as Ru$^{5+}$ and Os$^{5+}$, is less clear, as $L = 0$ in $L-S$ coupling.  The ordered moment for a number of Ru$^{5+}$ double perovskites is found to be $\sim 2$~$\mu_B$ rather than the spin only value of 3~$\mu_B$.\cite{Battle1989,Aharen2009} While SOC can contribute to such an effect, other factors may also be important as will be discussed later.  However, seemingly unequivocal experimental evidence for the effects of SOC is provided by inelastic neutron scattering studies of the material with which this study is concerned, $4d^3$ Ba$_2$YRuO$_6$.\cite{Carlo2013}.   A clear $\sim 5$~meV gap in the spin excitation spectrum, centered at the (100) magnetic reflection position is observed to develop below $T_{\rm N}$.  Such a gap is not expected for a pure $L = 0$ magnetic ground state, and we are very interested in whether a related spin gap occurs for $5d^3$ Ba$_2$YOsO$_6$.

One further unusual feature of some Ru$^{5+}$ double perovskites is the observation of at least two cases of two ordering temperatures, $T_{\rm N1}$ and $T_{\rm N2}$. For cubic Ba$_2$YRuO$_6$ these are 36~K and 47~K while for monoclinic Sr$_2$YRuO$_6$, 24~K and 32~K.\cite{Granado2013} For Sr$_2$YRuO$_6$, it has been argued that the region between $T_{\rm N1}$ and $T_{\rm N2}$ represents a mixed short range/long range ordering between AFM layers.\cite{Granado2013}.  Are multiple ordered states a characteristic of antiferromagnetic $d^3$ configurations on the FCC lattice and therefore also expected in $5d^3$ Ba$_2$YOsO$_6$?

In this paper we report structural and phase characterization including $^{89}$Y~NMR, bulk magnetic measurements, and elastic magnetic neutron diffraction, as well as spin dynamical characterization using time-of-flight neutron experiments on polycrystalline samples of Ba$_2$YOsO$_6$. This study shows obvious long range antiferromagnetic order below $T_{\rm N} \sim 69$~K, with magnetic Bragg peaks appearing at the [100] and [110] positions, consistent with a Type I antiferromagnetic structure with a reduced ordered moment on the Os$^{5+}$ site.  Neutron measurements of the order parameter show evidence for a second, weakly first-order transition just below $T_{\rm N}$, at $T_{c,2}= 67.45$~K.  The inelastic neutron data reveal a surprisingly large spin gap of $\Delta \sim 17$~meV develops below $T_{\rm N}$. These results are discussed in light of recent measurements performed on the related $4d^3$ double perovskite system Ba$_2$YRuO$_6$. 
%
\section{Experimental details and sample characterization}
%
Ba$_2$YOsO$_6$ was prepared by a solid state reaction according to the equation: 
\begin{equation} 
\text{Ba}_4\text{Y}_2\text{O}_7 + 2 \text{Os(10\% excess)} + \frac{5}{2} \text{O}_2 = 2 \text{Ba}_2\text{YOsO}_6
\end{equation}
The reaction was carried out in air at 1400\degree C for 1.5 hours in a Pt crucible with one regrinding after 30 minutes. The precursor material, Ba$_4$Y$_2$O$_7$, was prepared from stoichiometric amounts of BaCO$_3$ and Y$_2$O$_3$ fired at 1000\degree C in air for 24 hrs. Characterization by x-ray powder diffraction indicated a single phase sample, whereas powder neutron diffraction revealed a small amount ($\sim5\%$ in mass) of the weak ferromagnetic impurity Ba$_{11}$Os$_{4}$O$_{24}$. This does not interfere with our measurements on Ba$_2$YOsO$_6$ since this impurity orders at temperatures lower than, and distinct from, those relevant to ordering in Ba$_2$YOsO$_6$, i.e. 6.8~K.\cite{Wakeshima2005}

Magnetization measurements were performed using a Quantum Design MPMS magnetometer from 2~K to 300~K at various applied fields $\mu_0 H < 5$~T.
    
Neutron diffraction data were collected over the temperature range 3.5~K to 290~K using the C2 diffractometer at the Canadian Neutron Beam Centre, Chalk River, Ontario, Canada, using neutron wavelengths of 1.33~{\AA} and 2.37~{\AA} with no energy analysis. The sample, $\sim 2$g, was enclosed in a vanadium can in the presence of a He exchange gas. Crystal and magnetic structure refinements were performed using the FULLPROF analysis suite. Elastic neutron scattering experiments were also conducted on a $\sim 8$~g powder sample using the N5 triple-axis spectrometer at the Canadian Neutron Beam Centre, employing pyrolitic graphite (PG) as both monochromator and analyser, as well as a PG filter.  These measurements with a neutron wavelength of 2.37~{\AA} had a corresponding energy resolution of 1~meV (FWHM). The sample temperature was monitored using two sensors placed on either end of the sample can, showing a temperature difference of less than 0.1~K across top and bottom of the sample can between $T=65$~K and $T=70$~K.

$^{89}$Y~NMR was carried out at the University of Manitoba on a Bruker Avance III 500 ($B_0 = 11.7$~T) spectrometer operating at a Larmor frequency of 24.493~MHz. A 280~mg powder sample of Ba$_2$YOsO$_6$ was packed into a 4~mm (outer diameter) zirconia rotor and spun at 13.5~kHz. A set of 100k transients were averaged with a pulse repetition rate of 1.66~Hz. A rotor-synchronized Hahn-echo acquisition was used to eliminate probe ringing, with $\frac{\pi}{2}$ and $\pi$ pulses of 9 and 18~$\mu$s, respectively. Frequency shifts are referenced to 2M Y(NO$_3$)$_3$(aq) at 0 ppm. Due to the extreme temperature dependence of the paramagnetically shifted peaks, the sample temperature and its associated temperature gradient were calibrated using the $^{207}$Pb signal of solid Pb(NO$_3$)$_2$.\cite{Bielecki1995}

Inelastic neutron scattering measurements were performed at the Spallation Neutron Source (SNS, Oak Ridge National Laboratory), on the SEQUOIA Fine Resolution Fermi Chopper Spectrometer.\cite{Granroth2010} Incident neutron energies of 40 and 120~meV were employed, giving an energy resolution (FWHM) of 2.4 and 7~meV, respectively. The incident energies were chosen by Fermi chopper No.1 spinning at 180~Hz (40~meV) or at 300~Hz (120~meV). Background from the prompt pulse was removed by the T$_{\rm 0}$ chopper spinning at 90~Hz (40~meV) or 180~Hz (120~meV). 
The sample was enclosed in a planar aluminium cell with dimensions $5.0 \times 5.0 \times 0.1$~cm, under a He exchange gas atmosphere, and then loaded into an Orange $^4$He-flow cryostat, capable to achieve a temperature range of 1.5 to 300~K. An identical empty can was measured under the same experimental conditions and used for background subtraction. A boron nitride mask was used to reduce background scattering and normalization to a white-beam vanadium run was performed to correct for the detector efficiency differences.

\subsection{Crystal structure} 
\begin{figure*}
\includegraphics[width=1.7\columnwidth]{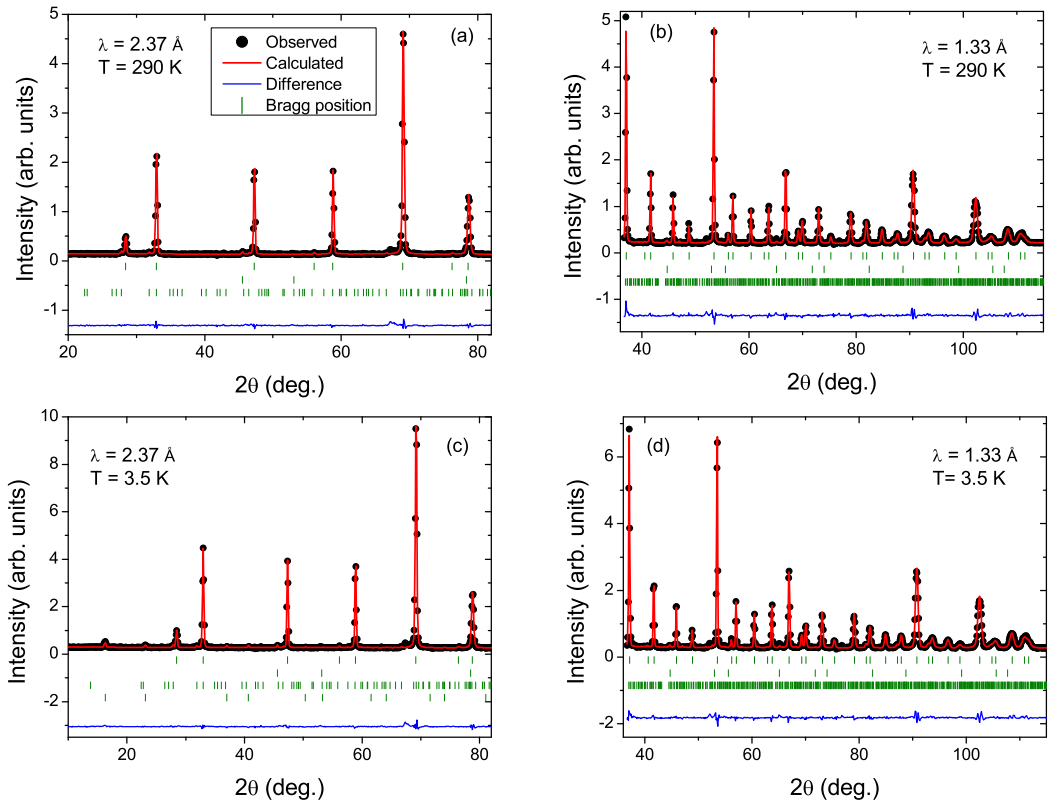}
\caption{\label{fig_diff} (Color online) Room temperature powder neutron diffraction patterns of Ba$_2$YOsO$_6$ recorded at 2.37~{\AA} (a) and 1.33~{\AA} (b). The 3.5~K neutron diffraction patterns at 2.37~{\AA} (c) and at 1.33~{\AA} (d) which shows that there is no structural change from 290~K. Tick marks represent Ba$_2$YOsO$_6$, Y$_2$O$_3$ ($\sim 1$\% in mass), and Ba$_{11}$Os$_4$O$_{24}$ ($\sim 5$\% in mass), respectively. }
\end{figure*}
The neutron powder diffraction data shown in Fig.\ref{fig_diff} was refined in space group $Fm\overline{3}m$, indicating a fully B-site ordered double perovskite structure at both 290~K and 3.5~K, indicating no change in symmetry over this range. The structural parameters resulting from these fits  are summarized in Table \ref{neutron_refin}. There is no evidence of $B/B'$ anti-site disorder from the neutron diffraction data.
\begin{table}
\caption{\label{neutron_refin} Neutron diffraction refinement results of Ba$_2$YOsO$_6$ at 290~K. The refinement was done simultaneously with two wavelengths, 2.37 {\AA} and 1.33 {\AA}. The results provided for the 1.33 {\AA} are reported in (~) behind the 2.37 {\AA} results and the 3.5~K refinement results are reported in [~] below the 290~K data.}
\begin{ruledtabular}
\begin{tabular}{cccccc}
Atoms         & x          & y &  z &   $B_{iso}$ (\AA$^2$)    \\
\hline
Ba          & 0.25 & 0.25 & 0.25 & 0.40(5) & \\
          &  &  &               & [0.01(5)] & \\
Y      & 0.50 & 0.50 & 0.50 & 0.38(9) & \\
          &  &  &               & [0.12(8)] & \\
Os      & 0 & 0 & 0 & 0.20 & \\
          &  &  &               & [0.15] & \\
O          & 0.235(2) & 0 & 0 & 0.64(4) & \\
          & [0.235(2)] &  &               & [0.29(4)] & \\
\end{tabular}
$a_0$ = 8.3541(4) {\AA} \\				
$\left[ a_0 = 8.3435(4) \mbox{{\AA}} \right]$ \\			
$R_{\rm Bragg}$ = 1.93 (3.26); $R_F$ = 1.41 (1.97); $\chi^2$ = 6.32 (10.0) \\
$\left[ R_{\rm Bragg} = 2.06 (2.63); R_F = 1.30 (1.66); \chi^2 = 7.04 (14.1)\right]$ 
\end{ruledtabular}
\end{table}
\subsection{Local Structure - Limits on $B/B'$ site disorder determined from $^{89}$Y NMR}
\begin{figure}
\includegraphics[width=0.8\columnwidth]{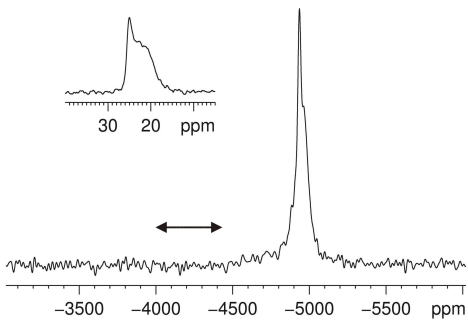}
\caption{\label{fig_NMR}  $^{89}$Y MAS NMR signal intensity of Ba$_2$YOsO$_6$ as a function of frequency ($B_0 = 11.7$~T) at 72\degree C.  (Inset: $^{207}$Pb MAS NMR of Pb(NO$_3$)$_2$ at 72\degree C.) The arrows define the region where a peak due to Y/Os anti-site disorder would be expected based on results from Ba$_2$YRuO$_6$.\cite{Aharen2009}}
\end{figure}

The extent of $B/B'$ anti-site disorder can be measured more accurately using magic angle spinning (MAS) NMR as demonstrated previously.\cite{Aharen2009} The results are displayed in Fig.\ref{fig_NMR} which shows a single peak in the NMR signal intensity as a function of frequency ($B_0 = 11.7$~T), centred near -4740~ppm. The low frequency shoulder is due to a thermal gradient associated with the sample spinning rotor as shown by the $^{207}$Pb reference signal of the inset. Any feature due to anti-site disorder would appear shifted by $\sim + 800$~ppm from the main peak as found to be the case in Ba$_2$YRuO$_6$.\cite{Aharen2009} We therefore conclude that any $B/B'$ anti-site disorder in Ba$_2$YOsO$_6$ is below the level of the noise.  An upper limit on the possible extent of $B/B'$ site disorder in our samples of Ba$_2$YOsO$_6$ is 0.5~\%, assessed as described in Ref.\onlinecite{Aharen2009}.

\subsection{Magnetization}
\begin{figure}
\includegraphics[width=0.85\columnwidth]{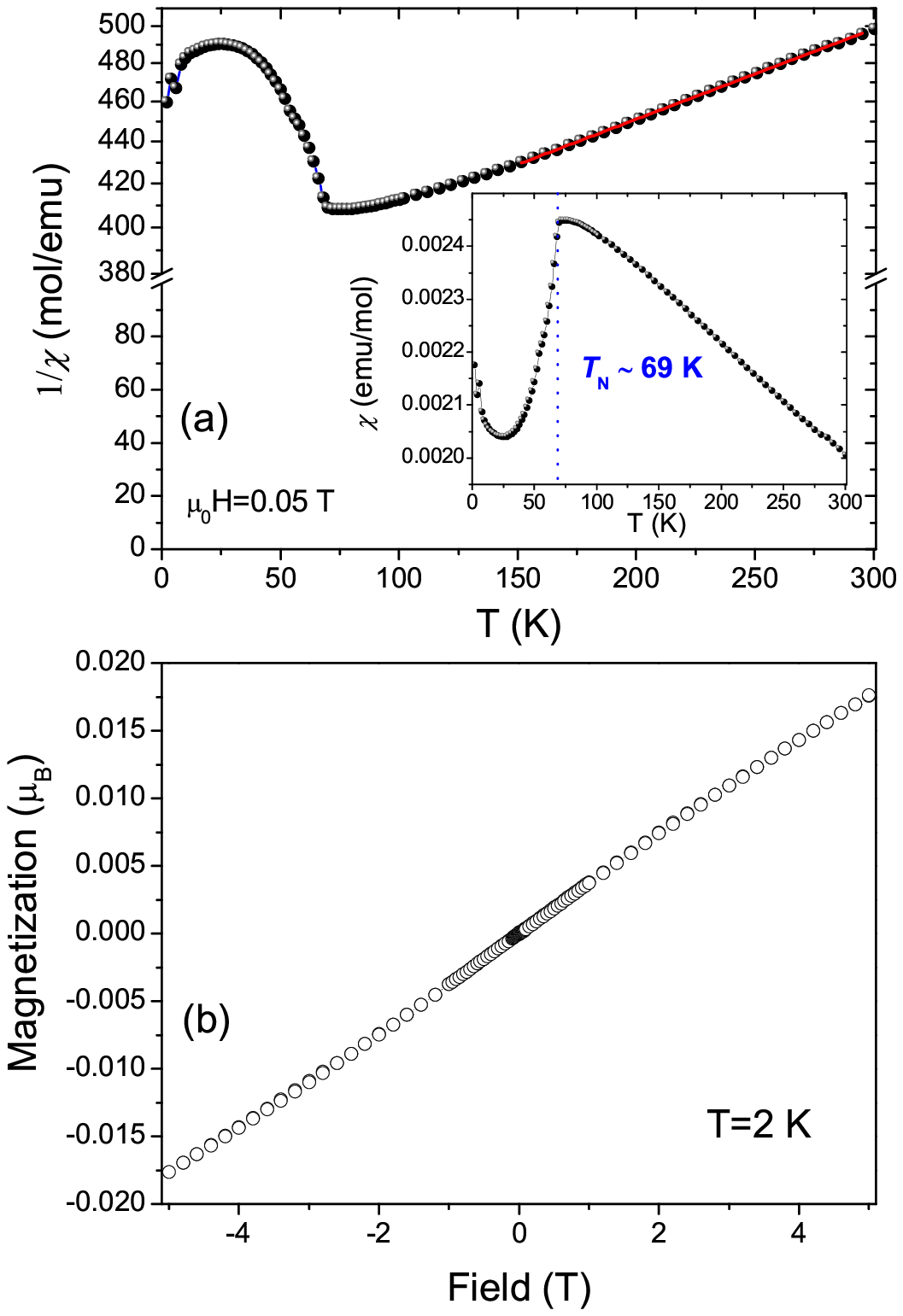}
\caption{\label{fig_mag}  (Color online) (a) The temperature dependence of the magnetic susceptibility (inset) and inverse magnetic susceptibility of Ba$_2$YOsO$_6$ at 0.05~T. The red line is a Curie-Weiss fit, performed for $150<T<300$~K. (b) The field dependence of magnetization at 2~K.}
\end{figure}

\begin{figure}
\includegraphics[width=0.85\columnwidth]{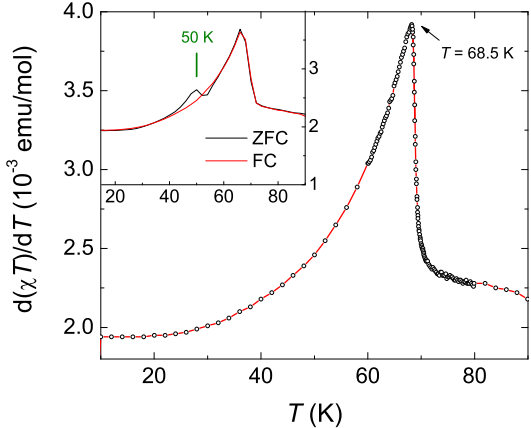}
\caption{\label{fig_Fischer} (Color online)  Fisher heat capacity of Ba$_2$YOsO$_6$. Main: a single peak is observable at 68.5~K, unlike the case of Ba$_2$YRuO$_6$. Inset: the peaks at $\sim 50$~K  can be attributed to the melting point of O$_2$. The O$_2$ peak is removed upon heating and is not observed in the FC curve.}
\end{figure}

Figure \ref{fig_mag}a shows $\chi(T)^{-1}$ (main panel) and $\chi(T)$ as measured in a magnetic field of 0.05 T.  We observe a clear anomaly in $\chi(T)$ at $T_{\rm N} \sim 69$~K, indicative of antiferromagnetic order.  In this $\chi(T)$ data there is no obvious evidence for multiple ordering temperatures, in contrast to the cases of $4d^3$ antiferromagnetic Ba$_2$YRuO$_6$ and Sr$_2$YRuO$_6$. A Curie-Weiss fit was attempted over the temperature range 150-300~K with the resulting parameters $\theta_{\rm cw} =  -772(4)$~K and a Curie constant $C =2.154(8)$~emu.K/mol, equivalent to an  effective moment $\mu_{\rm eff} = 4.151(8)$~$\mu_{\rm B}$. These parameters exceed the spin-only values relative to the Os$^{5+}$ moment, $C=1.87$~emu.K/mol and $\mu_{\rm eff} =3.87$~$\mu_{\rm B}$, by amounts far outside of the error bars of the fits.  This strongly suggests that the true paramagnetic regime is not found below 300~K. Nonetheless, the large negative $\theta_{\rm cw}$  reflects the presence of  strong antiferromagnetic interactions in this system and a frustration index $|\theta_{\rm cw}|/T_{\rm N} \sim 11$. No significant divergence between the zero field-cooled (ZFC) and field-cooled (FC) $\chi(T)$ curves is found for the entire temperature range measured. Furthermore, the field dependence of the magnetization at 2~K indicates no hysteresis and slight non-linearity for the dominant Ba$_2$YOsO$_6$ phase, implying that the ordered structure possesses no ferromagnetic component to it.  A plot of $d(\chi T)/dT$, the so-called Fisher heat capacity,\cite{Fisher1962} Fig.\ref{fig_Fischer}, shows a sharp lambda-like peak at 68.5~K. The weaker anomaly found near 50~K in the ZFC measurement (Fig.\ref{fig_Fischer}, inset) can be attributed to the presence of some solid O$_2$, removed upon heating. 
%
%

\section{Neutron scattering results}
%
\subsection{Magnetic Neutron Diffraction}
\begin{figure}
\includegraphics[width=0.9\columnwidth]{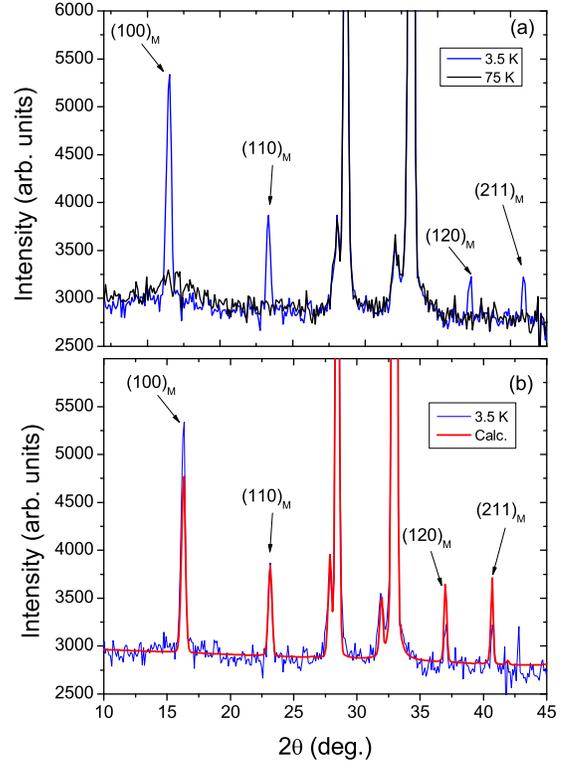}
\caption{\label{fig_elastic} (Color online) (a) Neutron diffraction patterns on Ba$_2$YOsO$_6$ collected at 3.5 and 75~K. The magnetic reflections are indexed with a propagation vector k = (1, 0, 0). The broad peak at the (100) magnetic reflection position indicates short-range magnetic correlations at 75~K. (b) Structural and magnetic refinements (red line) of the diffraction pattern observed at 3.5~K (blue line). The overfitting of the (120) and (211) magnetic reflections is ascribed to inaccuracies in the Os$^{5+}$ form factor.}
\end{figure}

Figure \ref{fig_elastic} shows a comparison of neutron diffraction data for Ba$_2$YOsO$_6$ at 75~K and 3~K taken with the C2 powder diffractometer. Four magnetic reflections are identified as they appear at low temperatures only. These peaks are indexed as indicated and are consistent with a Type I FCC  magnetic structure as found for several $4d^3$ Ru$^{5+}$ based double perovskites, including Ba$_2$YRuO$_6$.\cite{Battle1989,Aharen2009} The refined magnetic moment at the $5d^3$ Os$^{5+}$ site is 1.65(6)~$\mu_B$, significantly smaller than for Ru$^{5+}$ in Ba$_2$YRuO$_6$, 2.2(1)~$\mu_B$,\cite{Battle1989,Aharen2009} and for a number of other Ru$^{5+}$-based double perovskites which also show an ordered moment of 2~$\mu_B$.  The $5d^3$ Os$^{5+}$ ordered moment is also much reduced compared with the spin-only value of near 3~$\mu_B$ expected for the t$_{2g}^3$ electronic configuration.\cite{Battle2003}
%
A comparison between ordered moments found in perovskites with 4$d^3$, 5$d^3$ and 3$d^3$ electronic configurations is clearly of interest.  3$d^3$ Cr$^{3+}$, also t$_{2g}^3$, in the perovskite La$_{0.5}$Pr$_{0.5}$CrO$_3$ shows an ordered moment of 2.50(5)~$\mu_B$,\cite{Yoshii2001} confirming a progression to smaller ordered moments as SO coupling becomes progressively stronger.

What is the origin of the ordered moment reduction, of order $33-45$~\% between $4d^3$ Ba$_2$YRuO$_6$ and $5d^3$ Ba$_2$YOsO$_6$? Among  effects which can reduce the ordered moment are both SO coupling as well as covalency between the $d$ and ligand orbitals, which removes some fraction of the ordered moment from the transition metal ion site and transfers it to the ligand sites. Fluctuations induced by either geometrical frustration or quantum effects can also reduce the relevant ordered moments. However, given that Ba$_2$YRuO$_6$ and Ba$_2$YOsO$_6$ share the same frustrated FCC lattice and $d^3$ configuration, it is likely that SO coupling and covalency are more important effects.

The role of SO coupling is thought to be least important in the case where $B'$ has the t$_{2g}^3$ configuration, as to first order, $L = 0$, assuming $L-S$ coupling. The main effect is in second order which results in a reduction of the electronic $g$-factor by a term $\sim 8\lambda/10Dq$, where $\lambda$ is the SO coupling constant and $10Dq$ is the overall cubic crystal field splitting between the t$_{2g}$ and e$_g$ states.\cite{Abragam1970} In such a scenario, one expects that the ordered moment for t$_{2g}^3$ ions, $\mu = gS$, with $S = \frac{3}{2}$, will be somewhat reduced from the expected spin only value of 3~$\mu_B$. For Ba$_2$YRuO$_6$ and Ba$_2$YOsO$_6$, the SO coupling parameters for Ru$^{5+}$ and Os$^{5+}$ are estimated as $\lambda = 55$~meV and 186~meV, respectively.\cite{Ma2014} Taking  $10Dq$ in the range 2eV$-$3eV, the moments should fall between $2.7-2.8$~$\mu_B$ for Ru$^{5+}$ and $1.9-2.3$~$\mu_B$ for Os$^{5+}$.  Thus, SO coupling alone appears to be insufficient to fully explain the reduced ordered moments.

A similarly reduced ordered moment, 1.9~$\mu_B$, was found for the simple perovskite SrTcO$_3$ which involves the Tc$^{4+}$ ion, also 4$d$- t$_{2g}^3$.\cite{Rodriguez2011} In this case, it was argued from a first principles calculation that most of the moment reduction arises from 4$d$/ligand covalency.\cite{Mravlje2012,Middey2012} Nonetheless, very recently, it has been argued that SO coupling does play a major role in the magnetism of 4$d^3$ and 5$d^3$ perovskites in that a coupling scheme intermediate between $L-S$ and $j-j$ is required for a description of such a configuration.\cite{Matsuura2013} So, the role of SO coupling in 4$d^3$ and 5$d^3$ perovskite magnetic materials appears to be unsettled, at least regarding ordered moment reduction, at present.

\subsection{Neutron Order Parameter Measurements}
\begin{figure}
\includegraphics[width=0.9\columnwidth]{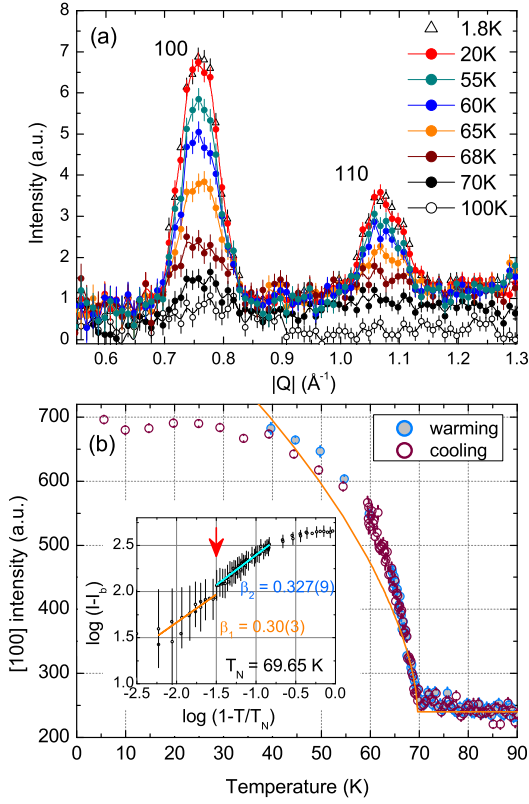}
\caption{\label{fig7_elastic} (Color online) (a) Temperature dependence of the elastic neutron scattering intensity in Ba$_2$YOsO$_6$, for $E_i = 40$~meV, integrated over the energy range $[-1,1]$~meV.  A long range order is observed below 70~K, with magnetic Bragg peaks at the $[100]$ ($|Q|=0.75$~{\AA}$^{-1}$) and $[110]$ ($|Q|=1.06$~{\AA}$^{-1}$) positions. A high temperature background dataset (150~K) has been subtracted from each data to isolate the magnetic contribution. (b) Temperature dependence of the [100] intensity from the N5 data (upon warming and upon cooling), showing a discontinuity at $T_{c,2} = 67.45$~K (red arrow). Blue and orange lines are fits to standard power-law to extract the critical behaviour (see text).} 
\end{figure}

We first discuss our elastic neutron scattering results derived from time-of-flight measurements, as shown in Fig.\ref{fig7_elastic}a. Measurements on Ba$_2$YOsO$_6$ with the incident energy $E_i = 40$~meV, and integrated over the elastic position [-1,1]~meV, show magnetic Bragg peaks at $|Q| = 0.75$~{\AA}$^{-1}$ and $|Q| = 1.06$~{\AA}$^{-1}$, corresponding to the $[100]$ and $[110]$ magnetic Bragg reflections, respectively (Fig.\ref{fig7_elastic}a. Their temperature-independent and resolution-limited widths in $|Q|$ stands as a clear signature of a 3D long range magnetic order, consistent with the type-I ordering of the FCC lattice near $T_{\rm N} \sim 69$~K, as observed from the C2 powder diffraction data presented above.

Additional elastic scattering measurements at the $[100]$ position, with enhanced temperature stability and much-finer temperature steps, were performed with the N5 triple-axis spectrometer at the CNBC, as shown in Fig.\ref{fig7_elastic}b.  The temperature of the sample was carefully equilibrated and stabilized to $\pm$ 0.005~K, and these measurements focused on the temperature dependence of the elastic magnetic intensity at the $[100]$ position.  The resulting order parameter, taken on both warming and cooling, for Ba$_2$YOsO$_6$ is shown in Fig.\ref{fig7_elastic}b.  These measurements show a small but clear discontinuity in the temperature dependence of the order parameter at $T_{c,2} = 67.45$~K, both on warming and cooling, indicative of a weak first-order transition below $T_{\rm N}$. The same data is plotted on a log-log scale in the reduced temperature, $(1-T/T_{\rm N})$ with $T_{\rm N}$=69.65 K, in the inset of Fig.\ref{fig7_elastic}.  A weak discontinuity in the shape of the order parameter manifests itself as a small change in the critical exponent $\beta$ estimated using data above and below $T_{c,2}$.  This is in contrast with Fisher heat capacity and neutron data reported above, which could only single out a unique transition around $T \sim 69$~K, but this could be due to the proximity of the two transition temperatures and the relative smallness of the discontinuity. 

The critical behaviour of the order parameter was fit using data above and below $T_{c,2}$, within the temperature ranges [59.4~K, 67.4~K] and [67.5~K, 69.2~K], to a standard power-law function:
\begin{equation}
I = A\left( \frac{T_{\rm N} - T}{T_{\rm N}}\right)^{2\beta_{i}} + I_b
\end{equation}
with $T_{\rm N}=69.65$~K and $I_b = 240$ as determined from a linear fit to the background intensity between 80 and 90~K, where the contribution from critical fluctuations is minimal. The two critical exponents are found to be $\beta_1 = 0.30(3)$ for $T_{\rm N}>T>T_{c,2}$ and $\beta_2 = 0.327(9)$ for $T<T_{c,2}$.  Both of these are consistent with 3D universality, and only $\beta_1 = 0.30(3)$ for $T_{\rm N}>T>T_{c,2}$ is meaningful as a critical exponent. Nonetheless, all of this critical and critical-like behaviour is consistent with a weakly first order transition at $T_{c,2} = 67.45$~K,  just below $T_{\rm N}$, and the presence of two different, ordered magnetic phases.

%

\subsection{Time-of-flight neutron spectroscopy and spin gap determination}

\begin{figure}
\includegraphics[width=\columnwidth]{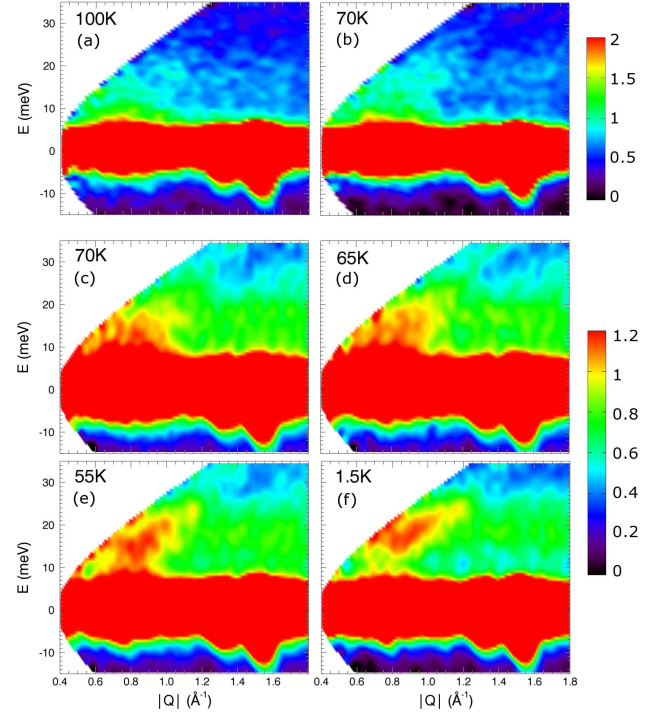}
\caption{\label{fig_Emaps} (Color online) (a)-(f) Evolution of the neutron scattering intensity $S(|Q|,E)$ across the magnetic transition at $T_{\rm N} \sim 69$~K. The background intensity of the sample cell has been subtracted off. Below 70~K, a spin gap $\Delta$ is progressively opening, finally reaching a $\sim 19$~meV value at 20~K and below. The upper intensity scale refers to (a) and (b) only, while the lower intensity scale refers to (c) to (e),  highlighting the opening of the gap.}
\end{figure}

We now turn to the description of our inelastic time-of-flight neutron measurements. The temperature variation of the neutron scattering intensity $S(|Q|,E)$ is shown in Fig.\ref{fig_Emaps}. The incident energy employed in these measurements, 120~meV, is high enough to allow an accessible region of $(|Q|,E)$ space sufficient for the observation of the full spin bandwidth, while being low enough such that the energy resolution of the measurement is suitable to access the inelastic gap. For $T \geq 70$~K, the spins are in a paramagnetic regime and we observe a quasi-elastic scattering mainly located at the $[100]$ position (Fig.\ref{fig_Emaps},a-b). As the temperature drops below $T_{\rm N}$, a clear gap of $\sim 17$~meV opens and separates the elastic (off-scale) intensity from the inelastic magnetic scattering, with a bandwidth presumably arising from the dispersion of the spin waves (c-f) within the ordered state. 

\begin{figure}
\includegraphics[width=\columnwidth]{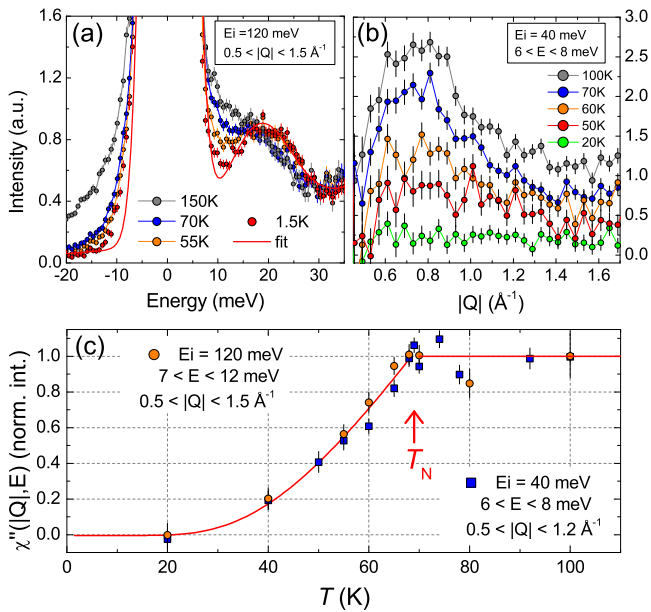}
\caption{\label{fig_inelastic} (Color online) (a) Energy cuts of Fig.\ref{fig_Emaps}, integrated over $Q=[0.5,1.5]$~{\AA}$^{-1}$, and showing the formation of the spin gap $\Delta$. The red line is a fit of the 1.8~K data to two-Gaussians centered on elastic and inelastic positions, resulting in a spin gap value of $\Delta = 19(2)$~meV and a spin wave bandwidth above the gap of 15(5)~meV. (b) Q cuts of the scattered intensity measured with an incident energy $E_i = 40$~meV, integrated over $E=[6,8]$~meV, i.e. in the inelastic regime, within the gap. The low $T$ dataset (1.8 K) has been subtracted as a background, and the plot shows the magnetic spectral weight filling the gap as the temperature is increased.  (c) $T$-evolution of $\chi''(|Q|,E)$, extracted from the data of panels (a) and (b) (see text), normalized by their respective plateau value above $T= 70$~K. Below 70~K, the intensity decreases exponentially (red line), allowing us to estimate $\Delta = 18(2)$~meV from this activated temperature dependence.}
\end{figure}

In order to better characterize the formation of the spin gap, we take different energy and wave-vector integrated cuts through both this dataset and the corresponding inelastic scattering dataset performed using $E_i = 40$~meV (see Fig.\ref{fig_inelastic}). The energy dependence of the low $|Q|$-integrated scattering  ($|Q| = [0.5,1.5]$~{\AA}$^{-1}$), mainly of magnetic origin, is shown in Fig.\ref{fig_inelastic}a for different temperatures. Upon cooling below 70~K, intensity within the gap region $7<E<12$~meV is depleted and transferred , at least in part, to higher energies where it contributes to the gapped and dispersive inelastic spin excitation. To quantify this, we fit the inelastic scattering at $T=1.8$~K in Fig.\ref{fig_inelastic}a to two Gaussians functions, one centered at each of the elastic and inelastic positions, along with a linear background to account for any weak inelastic background scattering, originating from, for example, phonons. The inelastic excitation is characterized by a spin gap $\Delta =18(2)$~meV and a bandwidth of 15(5)~meV, respectively defined by the center and the FWHM of the appropriate Gaussian. The temperature evolution of the spin gap can also be followed from the $|Q|$ dependence of the $E$-integrated scattering within the gap ($E = [6,8]$~meV), measured with $E_i = 40$~meV in order to obtain higher energy-resolution ($\Delta E = 2.4$~meV) (Fig.\ref{fig_inelastic}b). The low-$T$ dataset at 1.8~K offers an appropriate background subtraction, since minimal scattering is expected inside the gap. As the temperature is increased, we observed the progressive filling of the gap with a build-up of intensity near the $[100]$ region. The broad and asymmetrical lineshape observed in (Fig.\ref{fig_inelastic}b) is attributed to the presence of scattering coming up from the $[110]$ position, less intense than that near $[100]$ due to the appropriate structure factor and to its higher $|Q|$ value. 

With the background within the spin gap estimated using our base temperature data at $T=1.8$~K, we now examine the dynamical susceptibility $\chi''(|Q|,E)$, and eliminate the effect of the thermal Bose factor in the temperature dependence of the spin dynamics within the gap. We then obtain $\Delta S(|Q|,E)$,  which is related to $\chi''(|Q|,E)$ through:
\begin{equation}
\Delta S(|Q|,E) = \frac{\chi''(|Q|,E)}{1-e^{-E/k_{\rm B}T}}.
\end{equation}
The temperature behaviour of $\chi''(|Q|,E)$ can then be extracted from the $E$ and $Q$ integration of the scattering intensity of panels (a) and (b) in Fig.\ref{fig_inelastic}. The results, shown in panel (c), are in very good agreement and attest of the consistency of the measurements performed at two different incident energies. Below $T_{\rm N}$, the dynamic susceptibility shows an exponential behaviour, $\chi''(T) \propto \exp(-\Delta / k_{\rm B}T)$, consistent with a thermally activated excitation (red line, Fig.\ref{fig_inelastic}c). The spin gap extracted from this temperature dependence, $\Delta = 16(2)$~meV, is in good agreement with that previously estimated from the energy cut of panel (a). 

\section{Summary and conclusions}
The 5$d^3$ based double perovskite, Ba$_2$YOsO$_6$, has been prepared and characterized using a variety of methods including neutron and x-ray diffraction, $^{89}$Y NMR, magnetization and inelastic neutron scattering. To within the resolution of the elastic neutron diffraction data, the $Fm\overline{3}m$ space group is found for the entire temperature range to 3.5~K. The level of $B/B'$ anti-site disorder is below the level of detection from $^{89}$Y NMR studies ($<0.5$~\%). The material exhibits a strong antiferromagnetic Curie-Weiss constant, and magnetic frustration as characterized by a large frustration index, $f \sim 11$. Nonetheless, it displays at least one, and likely two phase transitions to ordered states resembling a Type I FCC antiferromagnetic ordered state below 70~K. 
While there is no obvious evidence from the Fisher heat capacity of two ordering temperatures, high resolution triple-axis elastic neutron scattering shows evidence for two transitions; a continuous one at $T_{\rm N} = 69.65$~K and a weakly first-order one at $T_{c,2} = 67.45$~K. This is reminiscent of the cases of the 4$d^3$ double perovskites Ba$_2$YRuO$_6$\cite{Carlo2013} and Sr$_2$YRuO$_6$,\cite{Granado2013} in which two ordering temperature scales, 36~K and 47~K or 24~K and 32~K respectively, were also apparent. The exact nature of the second ordered phase found in Ba$_2$YOsO$_6$, existing for $T_{c,2} < T < T_{\rm N}$, remains an open question. 

The low temperature ordered moment at the $5d^3$ Os$^{5+}$ site is 1.65(5)~$\mu_B$, much smaller than the spin only value of $\sim 3$~$\mu_B$.
It can be compared with low temperature ordered moments found in other $d^3$ configuration ions in perovskite structures, such as La$_{0.5}$Pr$_{0.5}$CrO$_3$ (2.50(5)~$\mu_B$) and Ba$_2$YRuO$_6$ (2.2~$\mu_B$).
Such a sequence of ordered moment reduction from 3$d^3$ to 4$d^3$ to $5d^3$ configuration suggests a strong quenching of the ordered moment from a combination of  SO coupling and transition metal-ligand covalency. A large spin gap, $\sim 17$~meV, is observed in the inelastic neutron scattering in $5d^3$ Ba$_2$YOsO$_6$ which can be compared with a spin gap of $\sim 5$~meV in 4$d^3$ Ba$_2$YRuO$_6$. As no spin gap is expected for the orbitally quenched t$_{2g}^3$ configuration, the spin gap is attributed to the effects of SO coupling. Indeed the ratio of the spin gaps for the $5d^3$ osmate and the 4$d^3$ ruthanate is $17/5 = 3.4$, essentially the same as the ratio of the free ion SO coupling constants for Os$^{5+}$ and Ru$^{5+}$, $186/55 \sim 3.4$,\cite{Ma2014} making a strong case for attributing the origin and magnitude of the spin gap to spin-orbit coupling.  

Recently, the low temperature properties of $5d^3$ distorted (monoclinic) double perovskite La$_2$NaOsO$_6$ was investigated with inelastic neutron scattering and $\mu$SR.\cite{Aczel2014} This system fails to display long range magnetic order at any temperature, presumably a consequence of its lower-than-cubic symmetry.  Below a spin freezing temperature of $T_f = 6$~K, the spins of this system enter into a static short-range incommensurate order and a spin gap never fully develops. These results bear some similarities with the orthorhombic $5d^3$ double perovskite Li$_3$Mg$_2$OsO$_6$,\cite{Nguyen2012} which also shows a frozen, short-range ordered ground-state, whereas the Ru analog orders at 17~K.\cite{Derakhshan2008} It would seem that the highly symmetrical FCC structure is a requirement for long-range magnetic order which in turn is  key to the formation of the spin gap.  Both Ba$_2$YOsO$_6$ and Ba$_2$YRuO$_6$ display a spin gap which forms as their magnetically ordered states form, below their appropriate $T_{\rm N}$. 

The combination of geometric frustration and SO coupling is a relatively unexplored arena in correlated electron physics.  5$d$ magnetism based on iridates has been of recent interest, but other 4 and 5$d$ systems involving osmates and ruthanates offer complex and exotic ground states, and are better suited to techniques such as neutron scattering.  Many related double perovskite structures can be stabilized, at least a subset of which display undistorted FCC lattices at low temperatures.   We anticipate that further basic and advanced characterization of these ground states will yield exciting new physics, as has already been predicted, for example, in $5d^1$ and $5d^2$ systems.\cite{Chen2010,Chen2011}

\section{Acknowledgements}

We acknowledge useful discussions with J. P. Clancy and thank J. P. Carlo for sharing information on experimental setup. SEQUOIA data were reduced using Mantid\cite{Mantid} and the DAVE software package\cite{DAVE} was used for data analysis. Research at Oak Ridge National Laboratory's Spallation Neutron Source was sponsored by the Scientific User Facilities Division, Office of Basic Energy Sciences, US Department of Energy. S.K. acknowledges the Canada Foundation for Innovation and Manitoba Research Innovation Fund for infrastructure support. S.K., J.E.G. and B.D.G. thanks the NSERC of Canada for operating funds via a Discovery Grant.

\end{document}